\documentclass{JHEP3}

\input{epsf}
\usepackage{epsfig}

\preprint{MIT-CTP-3396 \\
          OU-HET 450}

 \title{Towards black hole scattering}
\author{ 
Jan Troost and Asato Tsuchiya 
\\      Center for Theoretical Physics \\  MIT \\
    77 Mass Ave \\ Cambridge, MA 02139 \\ USA \\
\email{troost@mit.edu} \\ and \\ Department of Physics \\
Graduate School of Science \\
Osaka University \\ Toyonaka \\ Osaka 560-0043 \\ Japan 
\\ \email{tsuchiya@het.phys.sci.osaka-u.ac.jp}   }

 \abstract{We study black holes in three-dimensional
Chern-Simons gravity with a negative
cosmological constant. In particular,
we identify how the Chern-Simons interactions between
a scattering particle and a black hole project the particle wavefunction
onto a wavefunction in the black hole background. We also analyze the set
of space-times that should be allowed in the theory and the way in which boundary
conditions affect the spectrum of space-times.}

\begin{document}
\section{Introduction}
Three-dimensional gravity is a useful laboratory for studying physical
phenomena that are universal to theories of gravity in any dimension.
In particular, it is instructive to study three-dimensional gravity with 
a negative cosmological constant\cite{Deser:dr}, since it allows for non-trivial black hole
solutions \cite{Banados:wn}\cite{Banados:1992gq} (in contrast with three-dimensional gravity
with zero cosmological constant). Since a
lot of features of black hole physics are universal, it is reasonable to study
black holes in a simplified context.

The Einstein-Hilbert action in three dimensions, with a negative
cosmological constant can be rewritten in terms of Chern-Simons
theory with gauge group 
$SL(2,R) \times SL(2,R)$ \cite{Achucarro:vz}\cite{Witten:1988hc}.
In Chern-Simons theory,
we can treat these black hole excitations in roughly the same
 way as we would treat particles in three-dimensional gravity with zero
cosmological constant. Exact quantum scattering for 
particles in this context was studied  
in \cite{'tHooft:yr} in a metric approach and in \cite{Carlip:1989nz} in the
Chern-Simons formulation, with roughly equivalent results \cite{Carlip:1989nz}.
In the theory with negative cosmological constant, we can revisit
the problem, and try to study exact two-particle, two-black hole and black hole
- particle scattering. This study and some of its surprising aspects
are the subjects of our paper. But before we scatter sources we
discuss some intriguing aspects of three-dimensional gravity itself.

\section{Chern-Simons gravity}
\label{csgravity}
Three-dimensional
gravity with a negative cosmological constant
can be studied in terms of
 $SL(2,R) \times SL(2,R)$
Chern-Simons theory:\footnote{For our conventions and much more background on our
set-up, see \cite{Troost:2003ge}.}:
\begin{eqnarray}
S{[}A^+{]}_{CS}-S{[}A^-{]}_{CS} &=&  
  \frac{k}{4 \pi l} \int_M e_a \epsilon^{abc} R_{bc}
 + \frac{1}{3 l^2} \epsilon^{abc} e_a e_b e_c +
 \frac{k}{4 \pi l} \int_{\partial M} e^a \omega_a, \\
&=&  \frac{k}{4 \pi} \int_M Tr (A^+dA^+ +\frac{2}{3} {A^+}^3)
- \frac{k}{4 \pi} \int_M Tr (A^-dA^-+\frac{2}{3}{A^-}^3) \nonumber
\end{eqnarray}
where $A^{\pm} = \omega \pm \frac{e}{l} $ are two 
$SL(2,R)$ gauge fields and $e$ is the dreibein and $\omega$
is the spin connection one-form. The dimensionless 
constant $k=\frac{l}{4G}$ (where $G$ is
Newton's constant) measures
the radius of curvature associated to the cosmological constant
$\Lambda=-1/l^2$ in units
of the Plank length $l_p=G$. The  $SL(2,R) \times SL(2,R)$
Chern-Simons action is equal to the Einstein-Hilbert action up to boundary
terms. Our starting point for 
our discussion of the excitations in three-dimensional gravity will be the topological Chern-Simons theory on
a line times a disk $R \times D$, where we puncture the disk by including
a term in the action that couples a particle $\chi^{\pm}$ (that takes values in the
gauge group) minimally to the Chern-Simons gauge field \cite{Witten:1988hf}\cite{Witten:1989sx}.
We  briefly review the resulting picture of space-times.
The field strengths of the Chern-Simons theory must be thought of as the
curvature two-form and torsion two-form:
\begin{eqnarray}
\frac{F_a^+ + F^-_a}{2} &=& d \omega_a + \frac{1}{2} \epsilon_{abc} (\omega^b \omega^c + \frac{1}{l^2} e^b e^c)  \nonumber \\ 
\frac{F_a^+ - F^-_a}{2} &=& \frac{d e_a}{l} + \frac{1}{2l} \epsilon_{abc} (\omega^b e^c + e^b \omega^c).
\end{eqnarray}
In a certain gauge, and with an appropriate choice for the world-line time coordinate for the particle,
the equations of motion for the gauge fields $A_0^{\pm}$ imply \cite{Elitzur:1989nr}:
\begin{eqnarray}
\frac{F^+_{ij} + F^-_{ij}}{2} &=& -\frac{\pi}{k} \left( (\chi^{+} \lambda^{+} {\chi^{+}}^{-1})+
                                                   (\chi^{-} \lambda^{-} {\chi^{-}}^{-1})  \right) \delta \nonumber \\ 
\frac{F_{ij}^+ - F^-_{ij}}{2} &=&  -\frac{\pi}{k} \left( (\chi^{+} \lambda^{+} {\chi^{+}}^{-1})-
                                                   (\chi^{-} \lambda^{-} {\chi^{-}}^{-1})  \right) \delta,
\end{eqnarray}
where $\delta$ is a two-dimensional spatial delta-function, and $\lambda^{\pm}$ specify the orbits of the gauge
group on which 
the particle resides.
We note that when the particle source-term is zero, the geometry has constant curvature and zero torsion
(and negative cosmological constant $\Lambda=-1/l^2$). A solution to the equations of motion is given by
the $AdS_3$ geometry, which is gauge trivial. When we include a source-term
the topology becomes that of the real line times a disk $D$ with a puncture. We concentrate on sources
that are associated to particles that rest at the unit element of the gauge group 
($\chi^{\pm}=1$). It is clear that a particle will manifest itself as non-trivial
source terms for the curvature and torsion two-form. We argued in detail in  \cite{Troost:2003ge} that
we can associate hyperbolic weights $\lambda^{\pm}$ to generic BTZ space-times, a hyperbolic and a 
weight on the lightcone to extremal BTZ space-times and two lightcone weights to the massless BTZ black holes
\cite{Banados:wn}\cite{Banados:1992gq}.
In fact, we can add one more space-time to the list, which is the 
well-defined geometry (self-dual under T-duality) identified in \cite{Coussaert:1994tu},
which has trivial holonomy in one $SL(2,R)$ factor, and hyperbolic weight in the other.

We also want to identify the sources for conical space-times (without angular momentum) explicitly
(and more precisely than in \cite{Troost:2003ge}).
The torsion of these space-times will be zero, and the curvature singularity can be shown to 
be $R_{12} = 2 \pi \beta \delta^{(2)}$, 
where $2 \pi \beta$ denotes the deficit angle of the conical
geometry\footnote{This can be shown using the theory of distributions, or by transforming the conical
geometry to a conformally flat geometry with a singular conformal factor, and then regularizing
the conformal factor.}. These statements show that classical sources associated to conical space-times without
angular momentum have $\lambda^{+}=\lambda^{-}$ (because of zero torsion) and the weights
are elliptic: $\lambda^{\pm}
= k \beta  T_0^{\pm}$ (where $T_0$ generates an elliptic subgroup of $SL(2,R)$). 
Upon quantization, these classical source terms give rise (see \cite{Troost:2003ge})
to discrete particle representation spaces specified by the parameter $\tau^{\pm}=-\frac{k}{2} \beta -\frac{1}{2}$, when
$k>0$ and $\beta>0$.\footnote{The quadratic Casimir of a representation is given by $c_2=-\tau(\tau+1)$ in our
conventions.} In fact, we find that $\tau^{\pm}=-1$, which is the lowest lying true
discrete representation which can be obtained by quantizing orbits \cite{Vergne} is associated to a
space-time with conical deficit angle $2 \pi \beta= 2 \pi /k$. 
\subsection*{Mass}
With our definition for the bulk action in terms of Chern-Simons theory with no extra
boundary terms, 
we obtain a conformal field theory on the boundary which consists of two chiral $SL(2,R)$
Wess-Zumino-Witten models \cite{Elitzur:1989nr}. 
We propose to make use of the conformal symmetry of the boundary
theory to canonically define a mass operator. 
A natural definition of space-time mass $M_{CS}$ is given by the
time translation generator in the boundary conformal field theory, normalized in such a 
way that it satisfies the standard Virasoro algebra (in the quantum theory):
\begin{eqnarray}
L_0 &=& -\frac{\tau^+(\tau^+ +1)}{k-2} + osc \nonumber \\
\bar{L}_0 &=& -\frac{\tau^-(\tau^- +1)}{k-2} + osc \nonumber \\
M_{CS} & \equiv & L_0 + \bar{L}_0.
\end{eqnarray}
Black hole space-times correspond to continuous representations with quadratic Casimir
given in terms of the parameter $\tau^{\pm}=-\frac{1}{2}+is^{\pm}$
(and $s^{\pm} \in R$) and consequently have a positive mass $M_{CS}$ which is greater than
$M_{CS} = \frac{1}{2 (k-2)}$. The minimal mass which is associated to a CFT operator with positive
conformal dimension  is $M_{CS}=0$, which we obtain for $\tau^{\pm}=0$. It is the mass
of the $AdS_3$ space-time. We notice the counterintuitive fact that the space-time with conical deficit angle
$2 \pi/ k$ (associated to $\tau^{\pm}=-1$) also has zero space-time mass. Spaces with larger conical deficit
angle have a negative space-time mass $M_{CS}$.

Note that the mass gap (i.e. the gap in conformal dimension) between the $SL(2,R)$ invariant state and
the black hole continuum behaves as $1/k$ for large $k$ (i.e. in the
classical limit).\footnote{Note that the dimensionless ratio
$k=\frac{l}{4G}$ is the only coupling constant in the theory.} 
Quantum mechanically, there is a ``mass gap'' and complementary representations (with $-1<\tau<0$ and
$\tau \neq -1/2$) are excitations on
the vacuum that have mass smaller than the minimal black hole mass.

At this point we want to make remarks which clarify the relation between this theory of three-dimensional
gravity and the metric theory of three-dimensional gravity. It is known that  metric boundary
conditions (which insist on the fact that the metric asymptotes to $AdS_3$ (see e.g.\cite{Brown:nw})) gives rise to a Hamiltonian
reduction of the boundary $SL(2,R)$ conformal field theory. The Hamiltonian reduction is associated to a fixed
value for an $SL(2,R)$ current in a null-direction (see e.g.\cite{Polyakov:1987zb}\cite{Forgacs:ac}). 
It gives rise to a Liouville conformal field theory 
on the boundary\cite{Coussaert:1995zp}.
A natural definition of space-time mass in this metric theory is the sum of the left and right conformal dimensions,
as measured in the Liouville theory by the standardly normalized Virasoro zero-modes. This is the definition
adopted and analyzed in \cite{Krasnov:2002rn}. In this theory, the mass gap scales very differently. The
mass gap scales as $k$ with large $k$, in the classical limit, and the allowed space-times include the
full range of deficit angles from $0$ to $2 \pi$ \cite{Krasnov:2002rn}. 
It is clear then that these two theories of three-dimensional gravity 
are very different, and that the effect of the boundary conditions on the theory is very strong indeed.

We will further analyze the mapping between space-times and conformal field theory operators later on,
but we first turn to the scattering problem.


\section{Scattering two excitations}
We discuss the scattering of 
two  particles or black holes in the Chern-Simons formulation of
gravity with a negative cosmological constant. In three-dimensional
gravity with  zero cosmological constant,
the  papers \cite{'tHooft:yr}\cite{Carlip:1989nz} computed
the quantum amplitude for the scattering of two particles. 
(See also \cite{Carlip:1990mk}.) We
will follow \cite{Carlip:1989nz} closely in the following, and
we will generalize the analysis there to the case with 
negative cosmological constant. Since our results will closely
parallel those obtained in \cite{Carlip:1989nz}, we refrain from 
discussing many technical details. In fact, it is the 
conceptual assumptions that underlie the formalism which form the
trickiest part of the computation, so we discuss them in some detail.  
\subsection*{Base manifold scattering}
When we think of two-particle space-time scattering,
we usually think of a two-particle initial state at $t=-\infty$ and a two-particle final state
at $t=+\infty$. But we should realize that we not only have the base manifold concept of
time. Our Chern-Simons point-like particles also 
carry an ``internal clock'', because their internal wave-function is a Hilbert space
that is a representation of the $SL(2,R)\times SL(2,R)$ gauge group. 
For instance, wave-functions that
carry the same representation under the left and right $SL(2,R)$ gauge group can
simply be represented as wave-functions on the group manifold $SL(2,R)$. The Casimir
of the representation then specifies the (particle) mass of the wave-function, and the wave-function
in a coordinate representation satisfies a wave-equation with that mass. The wave-function
represents internal degrees of freedom, but because of the non-compact nature of the 
gauge group, includes a time-like variable. In that
sense each particle carries an ``internal clock''.

Ignoring this subtlety for a moment, we can compute the base manifold scattering amplitude of two particles
by specifying an initial state and a final state for the two-particle system, and then sum over all
paths with the appropriate phase. Since we work in a topological theory, we just have to sum
over all topologically distinct paths. These are enumerated by specifying the
number of times particle 1 winds around the world-line of particle 2. Summing the appropriate phases
will give the amplitude to scatter two 
particles.\footnote{We note that $\bigotimes$ which indicates a tensor-product of two one-particle states, should not
 be confused with $\otimes$ which will later indicate
 that a one-particle wave-function lives in a tensor
 product Hilbert space because of the product nature of the gauge group.}
We associate the following amplitude to this process:
\begin{eqnarray}
{\cal A} &=& 
\sum_{n=-\infty}^{\infty} \langle 1 | \bigotimes \langle 2 | B^n e^{i n \theta} |1 \rangle \bigotimes | 2 \rangle 
\end{eqnarray}
The operator $B$ is the braiding operation: it is the phase that the two-particle wave-function picks
up when particle 1 winds around particle 2 once (in the clockwise direction say) due to the Chern-Simons
interactions. 
The amplitude is a sum over all possible topological world-line histories, i.e. over all topologically
distinct paths. We have introduced an angle $\theta$ which specifies the phase which weighs the 
contributions to the path integral from the different topological sectors labeled by the winding number
$n$.
We will specify the precise form of the operator $B$ shortly.
\subsection*{Internal free motion}
Another quantity that we might be interested in is the following. Suppose we want to study a two-particle
state with particular initial conditions (at worldline time $t_1=-\infty$) 
for particle 1 and similarly (at worldine time $t_2=-\infty$)
for particle 2. There is dynamics in these time-variables simply because we know that the
wave-functions are $SL(2,R) \otimes SL(2,R)$ matrix elements. (E.g. for zero spin particles, the internal
dynamics is dictated by the Klein-Gordon equation with a mass squared given by minus 
the quadratic Casimir of the representation.) 
If we just study these wave-function by themselves, without referring to space-time, they evolve
freely. There is dynamics, but it is trivial.
The two particles do not interact as long as they do not move in space-time. They just ``sit'' in the base manifold
and evolve according to their internal clock.
\subsection*{Combination}
The way to reproduce a more intuitive concept of particle scattering in Chern-Simons theory, is to combine
the above concepts.
We study the wave-function for the two-particle system that evolves according to the internal clock,
but we also demand that we take into account the fact that the two-particle wave-function should move
from a given initial configuration to the same two-particle state at some final base manifold time $t$.
As a consequence, we know that the final state is a superposition 
\begin{eqnarray}
|\psi \rangle  &=& \sum_{n=-\infty}^{\infty} B^n e^{i n \theta} | \psi_0 \rangle
\end{eqnarray}
for some initial configuration $ | \psi_0 \rangle $
and that thus, it is invariant under the projection:
\begin{eqnarray}
e^{i \theta} B | \psi \rangle = | \psi \rangle.
\end{eqnarray}
This will be our definition of scattering: we evaluate the internal two-particle wave-function,
and project it out by the braiding operation, which represents the possibility of non-trivial
space-time topology for the particle world-lines.
This is the definition adopted in  \cite{Carlip:1989nz} and further analyzed in e.g.\cite{Koehler:px}\cite{deSousaGerbert:1990yp}
to which we refer for more details.


\section{Projection and periodicity}
After quantization, the source terms $J_i=\chi_i \lambda_i {\chi_i}^{-1}$
 for the Chern-Simons bulk gauge field act as currents $J^i$ on the quantum-mechanical
Hilbert space of the particles. The currents for each particle 
form an $SL(2,R) \times SL(2,R)$ algebra.
It has been argued in detail in  \cite{Carlip:1989nz} that the braiding operator
is given in terms of these current generators as:
\begin{eqnarray}
B &=& exp(\frac{2 \pi}{k} Tr(J_1 \bigotimes J_2)).
\end{eqnarray}
where the trace $Tr$ is over the full gauge algebra, and the lower indices indicate the
particle Hilbert space on which the currents act.
The same result for the braiding operator
 was obtained from a detailed analysis of open Wilson lines in Chern-Simons theory in
e.g.  \cite{Guadagnini:1989tj}.

We now have all the tools to analyze
the scattering of two excitations in the
Chern-Simons theory of gravity with negative cosmological constant. We will first discuss
the case where particle 2 can be treated as a classical source (i.e. we take the weights
$\lambda^{\pm}_2$ to be large compared to $\lambda^{\pm}_1$) that we fix to reside at the origin
of the gauge group (i.e. $\chi^{\pm}_2=1$ where $1$ is the unit element in the gauge group).

To analyze the action of the braiding/projection operator, we first discuss in more
detail the Hilbert space associated to excitation 1. We start out by sketching the familiar
picture for compact groups and then adapt the picture to our non-compact gauge group.
For compact groups $G$ the space of quadratically integrable functions decomposes into a sum of
tensor-products of irreducible Hilbert spaces:
\begin{eqnarray}
{\cal L}^2 (G) &=& \sum_{\lambda \, irr} {\cal H}_{\lambda} \otimes  {\cal H}_{\lambda}
\end{eqnarray}
 where the sum is over all irreps 
(labeled by a highest weight $\lambda$)
of $G$. For a non-compact group, a similar statement holds involving the 
representations that occur in the left/right regular representation and the summation
becomes an integral with a Plancherel
measure \cite{Vergne}. 

Note moreover that when we solve the Laplace/Klein-Gordon equation on the group manifold
in the space of quadratically integrable functions with given eigenvalue $-c_2(\lambda)$, then 
the solution space will span a representation of $G \otimes G$ which is 
${\cal H}_{\lambda} \otimes  {\cal H}_{\lambda}$. The wave-function with spin zero and mass squared
$-c_2 (\lambda)$ can be identified with a matrix element in the representation labeled
by $\lambda$, which is a vector in the ${\cal H}_{\lambda} \otimes  {\cal H}_{\lambda}$ representation
space.
We concentrate on a spin zero particle probe 1 and we can thus work with a wave-function
which is a solution to the Klein-Gordon equation, or in other words, a matrix element of an irrep.

\subsection*{Scattering off a black hole}

We are now ready to show how the Chern-Simons interaction mediates scattering off a target
black hole.
If we treat the target particle 2 classically, we can treat the associated currents
$J_2=\chi_2 \lambda_2 \chi_2^{-1}$ classically. If we assume that the target particle is at rest at the unit of the group
manifold, then we can equate the classical weight $\lambda_2$ with the current
$J_2$. For a black hole target, we have the classical weights $\lambda^{\pm}_2 = k \sqrt{M\pm J/l} T_1$
\cite{Troost:2003ge}.
Next, we parametrize the particle wave-function of particle 1 in terms of a function on the
group $G=SL(2,R)$. The parametrization we choose for the wave-function of particle 1 will be adapted
to the target particle 2. We parametrize the group manifold as $g=e^{ \frac{u}{2} \sigma_3} 
e^{ \rho \sigma_1} e^{\frac{v}{2} \sigma_3}$
where the coordinate transformation:
\begin{eqnarray}
\cosh^2 \rho &=& \frac{r^2-r_-^2}{r_+^2-r_-^2} \nonumber \\
u &=& \frac{r_+ - r_-}{l} (t+\phi) \nonumber \\
v &=& \frac{r_+ + r_-}{l} (\phi-t),
\end{eqnarray}
gives rise to the usual BTZ metric, but, most importantly, on the group the 
coordinate $\phi$ is {\em  not} identified modulo $2 \pi$.

The braiding projection operator can be represented in the
quantum Hilbert space of particle 1 by left and right multiplication of the argument
of the wave-function on the group manifold:
\begin{eqnarray}
B \psi_1(g) &=& e^{i \theta} \psi_1( e^{2 \pi i \sigma_3 (r_+ - r_-)/2l} g e^{2 \pi i \sigma_3 (r_+ + r_-)/2l}).
\end{eqnarray}
The simple computation yields an important result. When we have a spinless probe particle 1,
the projection condition
on the wave-function induced by the Chern-Simons interaction
is exactly the condition that the wave-function is periodic
in the coordinate $\phi$ with period $2 \pi$ (up to a possible phase given by
a $\theta$-angle). Thus the wave-function of particle 1 is interpreted as the wave-function
on a BTZ black hole background, {\em after} we implement the projection mediated by
the gravitational interactions. Note that this fact does not depend on the mass
of the spinless probe particle. In fact, from the algebraics of the geometry
of BTZ black holes \cite{Banados:1992gq}, and particle excitations
(see section \ref{csgravity}
and \cite{Troost:2003ge}), it is clear that the reasoning
holds quite generally for any type of probe or target particle.
(The literature  \cite{Ghoroku:1994ij}\cite{Natsuume:1996ij}\cite{Birmingham:2001dt}         
contains a detailed analysis of the resulting projected wave-functions in a different context.)
It would be very interesting to analyze the non-trivial dynamics that
occurs when both excitations, their currents, and their Hilbert spaces are treated quantum mechanically. We again stress that we explicitly showed that the 
quantum-gravitational scattering off a black hole allowed us to reconstruct the semi-classical
wavefunction for a particle in a black hole geometry.

\section{A broader picture}
\label{broader}
Although the probe-target approximation used in the previous section leads to 
intuitively plausible results for the scattering of a particle excitation of
a black hole, there are important counterintuitive features of the
formalism. These counterintuitive features can be seen to arise from the
following  observation. The space-time mass that we defined in section
\ref{csgravity} is, when we ignore boundary oscillatory excitations, proportional to minus the
mass squared that appears in the Klein-Gordon equation for the internal
(spinless) particle wave-function. More precisely, the correspondence between space-time mass and
KG mass squared is as follows. The space-time black hole mass spectrum corresponds to
KG masses that violate the Breitenlohner-Freedman bound. When we concentrate on particle excitations with
positive space-time mass, and which lie below the black hole mass spectrum,
we find that these have internal wave-functions corresponding to quadratic
Casimirs $0 \le \tau(\tau+1) \le -1/4$, which  implies that
they are ``stable tachyons'', that is, they have a negative mass squared which is above the Breitenlohner-Freedman
bound. When the space-time has negative mass $M_{CS}$, 
the internal KG mass squared is positive.

It is crucial to ask which space-times we should allow for in our theory of quantum gravity.
From the perspective of the CFT on the boundary, we may want to restrict to space-times with
positive mass (or, perhaps more appropriately put, positive conformal dimension). 
That demand excludes space-times with a deficit angle larger
than $2 \pi/k$. The only true discrete representations that would be allowed would
be $D_{-1}^{\pm}$, corresponding to $2 \pi/k$ deficit angle space-times.\footnote{It is 
intriguing to note that this naturally seems to give rise to $2k$ sectors in the multi-particle
Hilbert space, which was argued for on entirely different grounds in \cite{Fjelstad:2001je}.}
On the other hand, we know \cite{Vergne}\cite{Troost:2003ge} that the complementary representations
are not obtained from quantizing an orbit using the path integral method. We could nevertheless define
the quantum dynamics in these representations by algebraically extending (e.g. the braiding operation)
to the complementary representation Hilbert spaces. It is not clear whether we should
embrace these truly quantum Hilbert spaces that seem to have no corresponding classical geometry,
or whether we should just exclude them. From the boundary CFT point of view, certainly the continuous representations
are  acceptable and these correspond to the black hole space-times. They do lead to tachyonic
internal wave-functions. We note here on the one hand the clear analogy with Liouville theory (see e.g.\cite{Seiberg:1990eb}),
and on the other hand with the corresponding picture in three-dimensional gravity 
with zero cosmological constant (see e.g.\cite{Carroll:gc}).

We further remark that, since the tachyonic instability is clearly associated with the space-time itself 
(and not with the particle scattering off the space-time), the phenomenon of having a tachyonic wave-function
for a BTZ black hole is reminiscent of the instability of time-dependent
orbifolds (see e.g. \cite{Khoury:2001bz} and references thereto)
 in string theory (where one, as in the BTZ geometry,  also identifies space-time under a boost). 
There, in most scenarios, the space-time background itself is
unstable against collapse when a single excitation is added to the space-time. 
 Lastly, we mention the intriguing fact that it has been advocated recently
that space-like (i.e. tachyonic) geodesics can be used in the AdS/CFT context to probe regions of space-time behind
the event horizon of the black hole \cite{Kraus:2002iv}.

Note that we have often ignored oscillatory excitations on the boundary in the above.
We will start amending that omission in the following.

\section{Using CFT}
In this section we want to indicate another
approach to particle scattering which will allow us to make contact with another recent attempt to describe black hole
scattering in three-dimensional gravity \cite{Krasnov:2002rn}. It will consist of making use more heavily of the
connection between Chern-Simons theory on compact manifolds and two-dimensional conformal field theory \cite{Witten:1988hf}.
Suppose we have $n$ punctures in the disk which represent
the particles or black holes to be scattered.
The boundary of the disk represents the space at infinity. We
will need to specify boundary conditions to make the scattering
problem well-defined.
When we quantize the particle actions associated to the
punctures (by integrating over the particle degrees of freedom),
we obtain the expectation value of $n$ operators \cite{Witten:1988hf}:
\begin{eqnarray}
Z_n( R \times D) &=& \int dA e^{i S_{CS}} O_1 \dots O_n 
\end{eqnarray}
where the operators $O_i$ are given by \cite{Carlip:1989nz}
\begin{eqnarray}
O_i &=&
 {}_{\lambda^+_i \otimes \lambda^-_i} \langle \mbox{init} 
| P e^{\int A} | \mbox{fin} \rangle.
\end{eqnarray}
These are loop operators evaluated in the Hilbert space obtained
by quantizing the respective point particle action associated to the
weights $\lambda^+_i \otimes \lambda^-_i$. We need to specify an initial
and final condition for the particle path integral and these give rise to the evaluation
in an initial and final state. (We note that 
these open loop operators $O_i$ are gauge invariant since gauge transformations are
assumed trivial in the infinite past and future.)
\subsection*{Exact results}
At this point  we note that we can evaluate formal aspects of the scattering problem 
exactly by reasoning as follows. Suppose we compactify the time
in the base manifold, and moreover integrate over initial conditions
which we put equal to the final conditions (effectively tracing over the
Hilbert space). The resulting amplitude
to evaluate is:
\begin{eqnarray}
Z_n(S^1 \times D) &=& \int dA e^{i S_{CS}} W_1 \dots W_n
\end{eqnarray}
where the operators $W_i$ are now true Wilson loops in the
representations $ \lambda^+_i \otimes \lambda^-_i$.
Now, if we moreover specify a boundary conditions at infinity 
by inserting a Wilson loop on the boundary and gluing the disk
to an ``outer'' disk over their boundaries\footnote{We refer
to the case of a compact gauge group to argue for this prescription \cite{Witten:1989wf}.},
we obtain a spatial two-sphere $S^2$ with $n+1$ Wilson loop operators with a topology determined
by the scattering process under study:
\begin{eqnarray}
Z_n (S^1 \times S^2) &=& \int dA e^{i S_{CS}} W_1 \dots W_n W_{n+1}^{bc}.
\end{eqnarray}
Now, if we blindly analytically continue to the euclidean model and corresponding
 $SL(2,C)/SU(2)$ conformal field theory, we can use our knowledge of the conformal
field theory (\cite{Teschner:1997ft}) to evaluate these partition functions \cite{Witten:1988hf}.

We note that this type of formal analysis makes  contact with
the proposal in \cite{Krasnov:2002rn}, where Liouville amplitudes
(closely connected to $SL(2,C)/SU(2)$ amplitudes) are interpreted
as relevant to the scattering of black holes in the three-dimensional
theory of gravity with $AdS_3$ boundary conditions on the metric.
Of course, we have merely sketched the nature of the computation to be performed 
in this section. It would be interesting to flesh it out.

\section{Conclusions}
By applying known techniques in three-dimensional gravity to the case
with negative cosmological constant, we found that Chern-Simons interactions
can mediate gravity such as to reproduce scattering of particles off black holes.
At the same time we have shown that once the internal black hole wave-functions
is taken seriously, we run into interpretational difficulties (see discussion
in section \ref{broader}). 
We have argued that more information can be found
by carefully interpreting data of the conformal field theory on the boundary
as relevant to black hole scattering. We have stressed the crucial different behaviors
(as a function of the ratio of the cosmological radius and the Plank length)
of the theory with boundary conditions such that the metric asymptotes to $AdS_3$
and the theory with boundary conditions on the gauge connection consistent
with the full current algebra. We end by remarking that an AdS/CFT interpretation
of the Chern-Simons boundary conformal field theory would require copying the
 construction
of \cite{Giveon:1998ns}\cite{Kutasov:1999xu} for still another
space-time Virasoro algebra in the three-dimensional
gravity context. It will be interesting to see whether this third choice
of Virasoro algebra can help in resolving some of the counterintuitive features
that we found above, by implementing a more conventional picture of holography
(i.e. slightly closer to the string theoretic one \cite{Maldacena:1997re}).

\section{Acknowledgments}
Thanks to all the members of the CTP and especially
to Joe Minahan, Ashoke Sen, David Tong and Frank Wilczek for very useful discussions. 
A.T. is grateful to the CTP for its hospitality. Our
research was supported by the U.S. Department of Energy under cooperative 
research agreement \# DE-FC02-94ER40818.

\end{document}